\RequirePackage{graphicx}
\documentclass[aps,pra,floatfix,preprint, longbibliography]{revtex4-1}
\usepackage{amssymb}
\usepackage{enumitem}
\usepackage{mathtools}
\usepackage{xcolor}
\usepackage{float}
\usepackage{graphicx}
\usepackage{natbib}
\usepackage{dcolumn}
\usepackage{bm}
\usepackage{mathtools}
\usepackage{braket}
\usepackage[utf8x]{inputenc}
\usepackage{amsmath}
\usepackage{multirow}
\usepackage{amsfonts}
\usepackage{changepage}
\usepackage{rotating} 
\usepackage{amsthm}
\usepackage{tikz}
\usetikzlibrary{positioning}
\usetikzlibrary{patterns}
\begin{document}

%
\newcommand{\tc}{\textcolor{black}}

\title{\tc{Resolving Correlated States of Benzyne on a Quantum Computer with an Error-Mitigated Quantum Contracted Eigenvalue Solver}}

\author{Scott E. Smart, Jan-Niklas Boyn and David A. Mazziotti}
\email[]{damazz@uchicago.edu}
\affiliation{Department of Chemistry and The James Franck Institute, The University of Chicago, Chicago, IL 60637}
\date{Submitted March 26, 2021\tc{; Revised June 14, 2021}}

\begin{abstract}
The simulation of strongly correlated many-electron systems is one of the most promising applications for near-term quantum devices.
 \tc{Here we use a class of eigenvalue solvers (presented in \href{https://doi.org/10.1103/PhysRevLett.126.070504}{Phys. Rev. Lett. 126, 070504 (2021)}) in which a contraction of the Schr{\"o}dinger equation is solved for the two-electron reduced density matrix (2-RDM) to resolve the energy splittings of {\em ortho-}, {\em meta-}, and {\em para-}isomers of benzyne ${\textrm C_6} {\textrm H_4}$. In contrast to the traditional variational quantum eigensolver, the contracted quantum eigensolver solves an integration (or contraction) of the many-electron Schr{\"o}dinger equation onto the two-electron space.  The quantum solution of the anti-Hermitian part of the contracted Schr{\"o}dinger equation (qACSE) provides a scalable approach with variational parameters that has its foundations in 2-RDM theory. Experimentally, a variety of error mitigation strategies enable the calculation, including a linear shift in the 2-RDM targeting the iterative nature of the algorithm as well as a projection of the 2-RDM  onto the convex set of approximately $N$-representable 2-RDMs defined by the 2-positive (DQG) $N$-representability conditions. The relative energies exhibit single-digit millihartree errors, capturing a large part of the electron correlation energy, and the computed natural orbital occupations reflect the significant differences in the electron correlation of the isomers. }
\end{abstract}

\maketitle

\section{Introduction}

The simulation of many-body quantum systems is a key application for near-term quantum computing~\cite{Kassal2010,OMalley2016, McArdle2020, Arute2020}. The complexity of these simulations is such that algorithms on even moderately sized quantum devices---tens of qubits---with sufficient error mitigation will likely be competitive with existing classical methods~\cite{Aspuru-Guzik2005, Lloyd1996, Lu2012, Elfving2020}. A particular instance is the simulation of strongly correlated molecular systems, such as occur in many chemical reactions, transition-metal complexes, energetically degenerate processes, and solid-state materials~\cite{Helgaker2000, Lischka2018, Evangelista2018}. These systems, which often cannot be treated consistently with perturbative or polynomially scaling approaches relying on a single determinant, are ideal candidates for realizing an advantage from the use of quantum computers in lieu of classical computers, known as quantum advantage.  Realizing such as advantage, however, requires algorithms that are optimal for quantum computers in terms of state preparation, measurement, and error mitigation for the noise present in near-to-intermediate-term devices~\cite{Preskill2018, Head-Marsden2020}.


Various  variational quantum eigensolvers (VQE) for molecular simulation exist~\cite{Peruzzo2014, McClean2016, Kandala2017, Kandala2018, Fontana2020, McArdle2019, Smart2019hqca, Bonet-Monroig2018}, \tc{ most of which attempt to minimize the energy of a parameterizable ansatz against the Schr{\"o}dinger equation.} An alternative approach to electronic structure, rooted in reduced density matrix theory, involves a projection (or contraction) of the $N$-electron Schr{\"o}dinger equation onto the space of two electrons~\cite{Smart2020qacse_prl}, known as the contracted Schr{\"o}dinger equation (CSE)~\cite{Mazziotti1998b, Nakatsuji1996, Yasuda1997, Colmenero1993a, Valdemoro2008, Mazziotti2002d, Mazziotti1999a, Coleman2000}.  The CSE suggests an efficient ansatz and optimization strategy for computing the wave function. First, the solution of the CSE, it has been shown, produces an exact, rapidly convergent parametrization of the wave function from a product of only two-body exponential transformations~\cite{Mazziotti2004,Mazziotti2020}.  Furthermore, solution of the anti-Hermitian part of the CSE, known as the anti-Hermitian CSE (ACSE)~\cite{Mazziotti2006_prl, Mazziotti2007_pra, Gidofalvi2009, Mukherjee2001}, yields a parameterization of the wave function in terms of two-body unitary transformations~\cite{Mazziotti2006_prl, Gidofalvi2009}, which is theoretically exact~\cite{Evangelista2019} and readily implementable through unitary gates for state preparation on a quantum computer. Second, the residual of the ACSE yields the gradient of the energy with respect to two-body unitary transformations, which allows for more efficient optimization on quantum computers than derivative-free schemes~\cite{Peruzzo2014, Santagati2018, Robinson2006, Rakshit2018, Daskin2011} that could be limited to hundreds of parameters. \tc{Indeed, recent work by our group introduced a quantum algorithm which attempts to solve the ACSE using a quantum computer \cite{Smart2020qacse_prl}. }

Solution of the ACSE for the 2-RDM on classical computers has been applied to treating both ground and excited states of strongly correlated molecules including non-trivial conical intersections~\cite{Snyder2010, Snyder2011, Snyder2011c, Gidofalvi2009, Greenman2011, Sand2015, Alcoba2011, Smart2018}.  The solution of the ACSE on quantum computers, or quantum ACSE, can \tc{potentially} avoid the approximate reconstruction of the three-particle RDM (3-RDM) from the 2-RDM through preparation of the wave function on the quantum computer in polynomial time~\cite{Nielsen2010}. The quantum ACSE also shares certain similarities with the methods that attempt to decoupled and expand the single exponential unitary coupled cluster (UCC) ansatz~\cite{Romero2017, Lee2018}, such as the adaptive derivative-assembled pseudo-trotterization VQE (ADAPT-VQE) method~\cite{Grimsley2018}. \tc{The quantum ACSE circumvents issues of the \tc{trotterization of the ansatz (necessary for an exact exponential expression)} and high variational cost involved in an update step, and contains a natural selection of a pool of unitary transformations through the elements of the ACSE.} Moreover, because the ACSE generates the 2-RDM, it is readily combined with error mitigation strategies that correct the $N$-representability of the 2-RDM.  With its theoretical advantages and promising computational results, the ACSE method provides a potentially flexible framework for molecular simulation on quantum computers.

\tc{In the present work we apply the quantum ACSE solver to resolve the relative energies of the correlated isomers of benzyne on a superconducting quantum computer.}  The {\em ortho-}, {\em meta-}, and {\em para-}benzyne ($\textrm C_6 \textrm H_4 $) isomers contain non-trivial electron correlation, especially  {\em para-}benzyne which is a biradical~\cite{Yang2015, Shee2019, McManus2015, Wierschke1993, Debbert2000, Nash1996}.  The computed relative energies are accurate to less than 0.008~hartrees, and the natural-orbital occupations reflect the differences in electron correlation among the isomers.   The accuracy of the results demonstrates the benefits of both the solver and the error mitigation strategies.  Because these strategies are general, they can be applied to larger, more correlated molecules \tc{and represent a step towards performing strongly-correlated calculations on a quantum computer}

\section{Theory}

In this section we \tc{review} the theoretical framework for the quantum ACSE algorithm~\cite{Smart2020qacse_prl}, and \tc{explore the error mitigation schemes necessary for the calculation, including the use of $N$-representability conditions for the purification of the measured 2-RDM\cite{Foley2012,Rubin2018}}.

\subsection{Quantum Solver of the Anti-Hermitian Contracted Schr{\"o}dinger Equation}

For a many-electron system the two-electron contracted Schr{\"o}dinger equation~\cite{Mazziotti1998b, Nakatsuji1996, Yasuda1997, Colmenero1993a, Valdemoro2008, Mazziotti2002d, Mazziotti1999a, Coleman2000} is
\begin{equation}\label{CSE}
\langle \Psi | {\hat a}^\dagger_i {\hat a}^\dagger_j {\hat a}^{}_l {\hat a}^{}_k \hat{H}|\Psi \rangle = E ~{}^2 D^{ij}_{kl},
\end{equation}
where ${}^2 D$ is the 2-RDM, ${\hat a}^\dagger_i$ and ${\hat a}^{}_i$ are creation and annihilation operators for a spin orbital $i$, and $\hat{H}$ is the Hamiltonian operator that is given by
\begin{equation}
\hat{H} = \sum_{pqrs} {}^2 K^{pq}_{st} {\hat a}^\dagger_p {\hat a}^\dagger_q {\hat a}^{}_t {\hat a}^{}_s ,
\end{equation}
in which ${}^2 K $ is the reduced Hamiltonian matrix  containing the one- and two-electron integrals.  Taking the anti-Hermitian part of Eq.~\eqref{CSE} produces the ACSE~\cite{Mazziotti2006_prl, Mazziotti2007_pra, Gidofalvi2009, Sand2015, Alcoba2011}:
\begin{equation}\label{ACSE}
    \langle \Psi |  [{\hat a}^\dagger_i {\hat a}^\dagger_j {\hat a}^{}_l {\hat a}^{}_k ,\hat{H}]|\Psi \rangle = 0 ,
\end{equation}
which depends upon not only the 2-RDM but also the 3-electron RDM (3-RDM) (see Refs.~\cite{Mazziotti1998, Mazziotti2007_pra, Deprince2007} and Appendix).  The residual of the ACSE is equal to the gradient of the energy with respect to two-body unitary transformations, and hence, the residual of the ACSE vanishes if and only if the gradient vanishes.  Consequently, the ACSE provides a framework for the iterative optimization of a product of two-body unitary transformations on a reference wave function, \tc{which leads to the quantum ACSE algorithm presented in Figure 1.}

Let $|\Psi_n\rangle$ be the $n$-th iteration of the wave function, where  ${}^2 D_0$ is the 2-RDM of the initial Hartree-Fock state $|\Psi_0\rangle$. The 2-RDM of the $(n+1)$-th iteration is
\begin{equation}\label{ansatz}
{}^2 D^{pq;st}_{n+1}  = \langle \Psi_n | e^{-\epsilon_n \hat{A}_n} {\hat a}^\dagger_p {\hat a}^\dagger_q {\hat a}^{}_t {\hat a}^{}_s  e^{\epsilon_n \hat{A}_n}|\Psi_n  \rangle,
\end{equation}
where $\epsilon_n$ is theoretically an infinitesimal step and ${}^2 {\hat A}_n$ is an anti-Hermitian operator
\begin{equation}
    {}^2 {\hat A}_n = \sum_{ijkl} {}^2 A_n^{ij:kl} {\hat a}^\dagger_i {\hat a}^\dagger_j {\hat a}^{}_l {\hat a}^{}_k.
\end{equation}
The energy at each iteration is computable from the 2-RDM
\begin{equation}
E_{n+1} = \sum_{pqst}{{}^{2}K^{pq}_{st} ~^{2}D^{pq;st}_{n+1} } .
\end{equation}
Elements of the ${}^2 A_n$ matrix can be selected~\cite{Mazziotti2007_pra} as the residual of the ACSE
\begin{align}
    {}^2 A_n^{ij;kl} = \langle \Psi_n | [{\hat a}^\dagger_i {\hat a}^\dagger_j {\hat a}^{}_l {\hat a}^{}_k,\hat{H}]| \Psi_n \rangle,
\end{align}
which is effective because the ACSE's residual is related to the gradient of the energy with respect to the elements of ${}^2 A_n$
\begin{align}\label{classA}
     \langle \Psi_n | [{\hat a}^\dagger_i {\hat a}^\dagger_j {\hat a}^{}_l {\hat a}^{}_k,\hat{H}]| \Psi_n \rangle = -\frac{1}{\epsilon_n} \frac{\partial E_{n+1} }{\partial ~{}^2 A_n^{ij;kl}} + O(\epsilon_n).
\end{align}
Hence, by using the residual, we are choosing a search direction that maximizes the change in the energy for small $\epsilon_{n}$.  The ACSE can be expressed in terms of the 2- and 3-RDMs and can be evaluated classically with an $O(r^6)$ cost using a reconstructed 3-RDM in which $r$ is the rank of the one-electron basis set.  On a quantum computer, we can obtain elements of ${}^2 A_n$ in a potentially more efficient manner without the reconstructed 3-RDM.  Define an auxiliary 2-RDM:
\begin{equation}
    {}^2_{\pm} \Lambda_{n}^{ij;kl}  = \langle \Psi_n | e^{\mp i\delta\hat{H}} {\hat a}^\dagger_i {\hat a}^\dagger_j {\hat a}^{}_l {\hat a}^{}_k  e^{\pm i\delta \hat{H}}|\Psi_n\rangle ,
\end{equation}
in which the $n$-th wave function is propagated through a time-like step $\delta$ in the forward or reverse direction. Then, we can obtain elements of the residual from tomography of these auxiliary RDMs with $O(r^4)$ scaling:
    \begin{align}\label{quantA}
     {}^2 A_n^{ij;kl} =\frac{1}{2 i\delta}({}^2_+ \Lambda^{ij;kl}_{n} - {}^2_- \Lambda_n^{ij;kl}) + O(\delta^2).
\end{align}
These equations suggest an iterative approach to finding a solution of the ACSE, which is depicted in Fig.~1. After initializing the wave function and 2-RDM, for a given iteration we construct the operator ${}^2 A$ through classical or quantum approaches, prepare and measure $ {}^2 D_{n+1}$ (possibly optimizing $\epsilon_n$ and carefully selecting elements of ${}^2A_n$ to include in the wave function), and iterate between ${}^2 D_{n}$ and ${}^2 A_{n}$ until $|| {}^2 A_n||$ is less than a certain threshold.

In the classical-computing algorithm the solution of the ACSE requires an approximate reconstruction of the 3-RDM from the 2-RDM through a cumulant expansion~\cite{Mazziotti1998, Kutzelnigg1999, Misiewicz2020} to compute the 2-RDM without the wave function.  In the quantum-computing algorithm, in contrast, the wave function is prepared with polynomial scaling, and hence, approximate reconstruction of the 3-RDM is not necessary.  In the noiseless limit the ACSE can be solved by the quantum-computing algorithm to an arbitrary level of accuracy.  The errors arising from the expansion in Eq.~(6) are controllable with respect to $\delta$.  Computationally, we find in the noiseless limit that the solution of the ACSE yields a wave function, parameterized by two-body unitary transformations, that solves not only the ACSE but also the $N$-electron Schr{\"o}dinger equation.

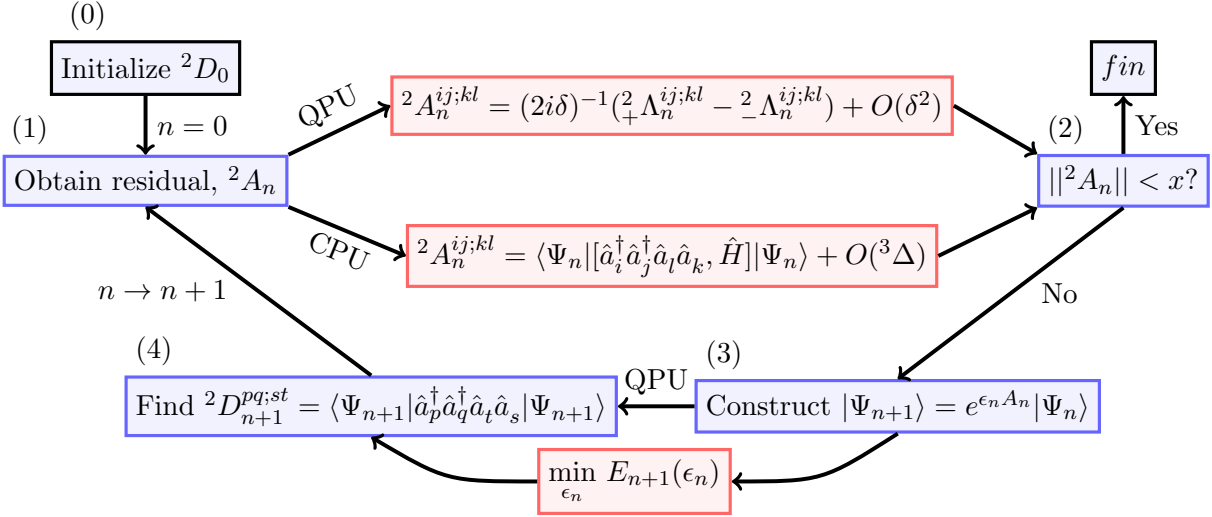
\begin{figure}
\caption{The quantum-ACSE algorithm. After initializing the state (0), we begin an iterative process of obtaining the ${}^2 A$ elements (1) through a quantum (QPU) or classical (CPU) processor, which will have errors in the series expansion ($O(\delta^2)$) or 3-electron reduced cumulant matrix (${}^3 \Delta$), respectively. After checking for convergence against a threshold $x$ (2), we construct the next ansatz (3), and optionally perform a classical minimization against the step size. Finally, we measure the new 2-RDM (4), and proceed to (1) until we converge or $n=n_{\rm max}$.}
\begin{tikzpicture}[
round/.style={rectangle, draw=black, fill=blue!5, very thick, minimum size=7mm},
square1/.style={rectangle, draw=red!60, fill=red!5, very thick, minimum size=5mm},
square2/.style={rectangle, draw=green!60, fill=green!5, very thick, minimum size=5mm},
square3/.style={rectangle, draw=blue!60, fill=blue!5, very thick, minimum size=5mm},
blanknode/.style={circle},
]
\node[round,label={[xshift=-2em] (0)}] (t0) at (-3,2) {Initialize ${}^2 D_0$};
\node[square3,label={[xshift=-4em] (1)}]   (t1) at  (-3,0.5)    { Obtain residual, ${}^2 A_n$};
\node[square1]   (c) at (4,-0.5) {${}^2 A_n^{ij;kl} = \langle \Psi_n | [{\hat a}^\dagger_i {\hat a}^\dagger_j {\hat a}^{}_l {\hat a}^{}_k,\hat{H}]| \Psi_n \rangle + O({}^3 \Delta) $};
\node[square1]   (q) at (4,1.5) {${}^2 A_n^{ij;kl}  = (2 i\delta)^{-1}({}^2_+ \Lambda^{ij;kl}_{n} - {}^2_- \Lambda_n^{ij;kl}) + O(\delta^2)$};
\node[square3,label={[xshift=-2em] (2)}]   (t2) at (10,0.5) { $||{}^2 A_n || < x ?$};
\node[square3,label={[xshift=-6em] (3)}]   (t3) at  (7,-2.5) {Construct $| \Psi_{n+1}\rangle = e^{ \epsilon_n A_n} |\Psi_n \rangle$};
\node[square3, label={[xshift=-7.5em] (4)}]   (t4) at (0,-2.5) {Find ${}^2 D_{n+1}^{pq;st} = \langle \Psi_{n+1} |{\hat a}^\dagger_p {\hat a}^\dagger_q {\hat a}^{}_t {\hat a}^{}_s | \Psi_{n+1} \rangle $};
\node[square1]   (t5) at (3.5,-3.5) {\small $\min\limits_{\epsilon_n}$ $E_{n+1}(\epsilon_n)$ };
\node[round]  (f) at (10,2) {$fin$};

\draw[->,line width=1.5pt] (t0.south) -- node[right]{$n=0$} (t1.north);
\draw[->,line width=1.5pt] (t1.south east) -- node[below,sloped]{CPU} (c.west) ;
\draw[->,line width=1.5pt] (t1.north east) -- node[above,sloped]{QPU} (q.west);
\draw[->,line width=1.5pt] (c.east) -- (t2.south west);
\draw[->,line width=1.5pt] (q.east) -- (t2.north west);
\draw[->,line width=1.5pt] (t2.south) -- node[right]{~~No} (t3.north);
\draw[->,line width=1.5pt] (t3.west) -- node[above]{QPU} (t4.east);
\draw[->,line width=1.5pt] (t4.north) -- node[left]{$n\rightarrow n+1$~~~}(t1.south);
\draw[->,line width=1.5pt] (t2.north) -- node[right]{Yes} (f.south);

\draw[->,line width=1.5pt] (t3.south) .. controls (6,-3.5) .. (t5.east);
\draw[->,line width=1.5pt] (t5.west) .. controls (1,-3.5) .. (t4.south);

\end{tikzpicture}
\end{figure}

Finally, several variations of the algorithm are possible for practical implementations on quantum computers. For example, a limited portion of ${}^2 A_n$, such as its largest terms, can be used; a stochastic gradient or reduced gradient sampling technique can be implemented, lowering the measurement cost of ${}^2\Lambda_n$ at each step. The quantum and classical methods can 1be combined where direct quantum tomography is only employed for the parts of the 2-RDM that are strongly correlated.

\subsection{Quantum Computation}
In this work we utilize the qACSE method and generate ${}^2 D_n$ on the quantum computer, and obtain elements of ${}^2 A_n$ on the quantum computer for the smaller qubit calculations (Eq.~\ref{quantA}), and classically with a reconstructed 3-RDM for the larger qubit calculations (Eq.~\ref{classA}). Figure 2 provides an overview of the process to obtain a fully error mitigated ${}^2 D_n$. We also include details related to the specific techniques and other aspects of the calculation in the Appendix and Supplemental Material.

To obtain ${}^2 D_n$, at a given step, we first transform the ${}^2 A_n$ operator into a suitable form for the quantum computer (including our qubit reduction scheme). \tc{Explicitly, this is done through a first order trotterization of the exponential of Eq. (5), where each element of the ${}^2 A_n$ matrix is implemented separately. However, because we would like to avoid implementing all the operators at once, we use an element threshold to determine inclusion in the ansatz. To implement the gate sequence, we prepare and manually simplify the set of 2-RDM operators corresponding with possible elements of ${}^2 A_n$. These are assembled according to our inclusion criteria, and then the circuits are run}. After measurement, we apply a filter (via construction and inversion of a state transition matrix, referred to as SPAM) and then apply a projection into the proper number and projected spin space ($N\in \{2,4 \}$, $S_z=0$) for measurements which commuted with these operators (which are $Z_i$ type measurements). In some cases we then apply our limit-preserving correction $\Gamma_n$ to the ansatz (see below), followed by an optional purification of the $2$-RDM.

\begin{figure}\caption{Error mitigation scheme to obtain corrected 2-RDMs. We first take a set of instructions, and construct the appropriate circuit design. We run these on the quantum computer to obtain a set of measurement results which are then \tc{corrected through the inversion of a state preparation matrix} (SPAM, small hatched rectangle). Measurements corresponding to diagonal elements of the 2-RDM ($M_i^c$) will commute with the $\hat{N}$ and $\hat{S}_z$ symmetries, and so are projected onto the proper operator space. We then apply our shift correction, $\Gamma_n$, which also preserves trace but can introduce negative eigenvalues, and optionally, a purification of the 2-RDM.}
\begin{tikzpicture}[round/.style={rectangle, draw=black, fill=blue!5, very thick, minimum size=7mm},
square1/.style={rectangle, draw=black!60, fill=red!5, very thick, minimum size=5mm,
},
square2/.style={rectangle, minimum size=5mm},
square3/.style={rectangle, draw=blue!60, fill=blue!5, very thick, minimum size=5em,
text width=10em, minimum height=10em},
blanknode/.style={circle},
]

\draw[blue,fill=blue!5] (1,-2.2) rectangle (4.8,2.0) node[label={[xshift=-8.5em,yshift=-2em] QPU}]{};
\node[square1] (s0) at (-1,-1.3) {Ideal $|\Psi_n\rangle$};
\node[square1] (s1) at (-1,1) {$ U_n |00..\rangle$};
\node[square2] (s2a) at (3,1) {$ \langle 00..| U^\dagger_n M_i^c U_n |00..\rangle$};
\node[square2] (s2b) at (3,0.33) {$ \langle 00..| U^\dagger_n M_j^c U_n |00..\rangle$};
\node[square2] (s2c) at (3,-0.33) {$ \cdot \cdot \cdot $};
\node[square2] (s2d) at (3,-1) {$ \langle 00..| U^\dagger_n M_i U_n |00..\rangle$};
\node[square2] (s2e) at (3,-1.67) {$ \langle 00..| U^\dagger_n M_{j} U_n |00..\rangle$};

\node[square1] (s4) at (6.5,-1.33) {\large ${}^2 D_n $};
\node[square1] (s5) at (9,-1.33) {\large ${}^2\tilde{D}_n $};
\node[square1] (s6) at (9,1) {$P({}^2 \tilde{D}_n )$};

\node[square2] (text1) at (6,-2.3) {SPAM};

\draw[pattern=north west lines, pattern color=black] (4.9,-2.2) rectangle (5.0,2);
\draw[red,fill=red!5] (5.1,-0.33) rectangle (5.85,2) node[label={[xshift=-0.35cm,yshift=-2em]$P_N$}]{} node[label={[xshift=-0.35cm,yshift=-3.3em]$P_{S_z}$}]{};
\draw[->] (s0) -- node[left]{Compilation} (s1) ;
\draw[->] (s4) -- node[above]{$+\Gamma_n$}(s5);
\draw[dashed,->] (s5) -- node[right]{Purify}(s6);

\draw[->,line width=1.5pt] (s1.east) .. controls (0.2,1) and (0.5,1) .. (s2a.west);
\draw[->,line width=1.5pt] (s1.east) .. controls (0.2,1) and (0.5,0.33) .. (s2b.west);
\draw[->,line width=1.5pt] (s1.east) .. controls (0.2,1) and (0.5,-1) .. (s2d.west);
\draw[->,line width=1.5pt] (s1.east) .. controls (0.2,1) and (0.5,-1) .. (s2e.west);

\draw[->,line width=1.5pt] (s2a.east) .. controls (5.5,1) and (6.5,0.33) .. (s4.north);
\draw[->,line width=1.5pt] (s2b.east) .. controls (5.5,0.33) and (6.5,0.33) .. (s4.north);

\draw[->,line width=1.5pt] (s2d.east) .. controls (5.5,-1.0)  .. (s4.west);
\draw[->,line width=1.5pt] (s2e.east) .. controls (5.5,-1.66)  .. (s4.west);

\draw[->,line width=1.0pt] (text1.west) -- (5+0.05,-2.2);

\end{tikzpicture}
\end{figure}
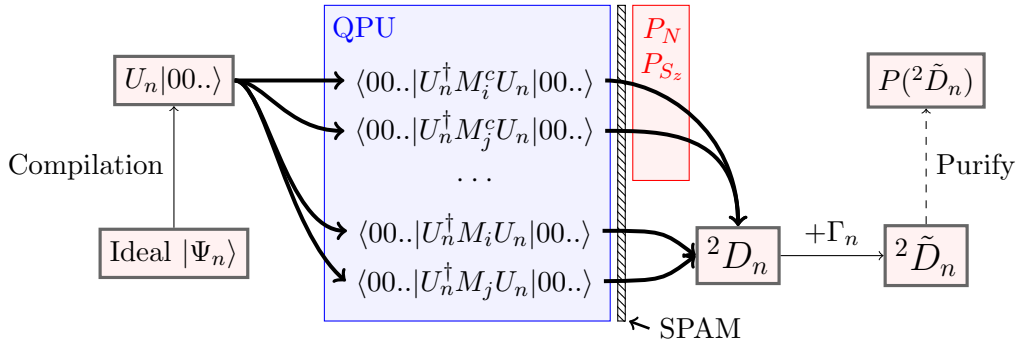

To our knowledge, the \tc{penultimate} error mitigation technique has not been used elsewhere, \tc{and the final technique was recently introduced for quantum simulations \cite{Rubin2018} but not yet demonstrated experimentally}, and so we briefly detail them here. The first is a correction targeting errors in an iterative ansatz that arise simply from adding extra gates, whereas the second is an expansion of techniques related to ensuring the physicality of the measured RDM through $N-$representability constraints.

\subsection{Limit-Preserving Correction for an Iterative Ansatz}

To compensate for errors which occur at each step due to the increasing the number of gates in an iterative scheme, we present an error mitigation strategy which we call a limit-preserving correction or a ${}^2\Gamma$-correction. Consider the $n$-th iteration of the qACSE algorithm. Given the elements of ${}^2A_n$, we can consider the $(n+1)$-th 2-RDM as a function of $\epsilon_n$ as it approaches $0$ from the positive direction:
\begin{align}
    {}^2 D^{pq;st}_{n+1}(0^+) &= \lim_{\epsilon_n \rightarrow 0^+} {}^2 D^{pq;st}_{n+1}(\epsilon_n) \\
    &=   {}^2 D^{pq;st}_{n}(\epsilon_{n-1}) + \lim_{\epsilon_n \rightarrow 0^+}  \epsilon_n \langle \Psi_n |[a^\dagger_p a^\dagger_q a^{}_t a^{}_s,{}^2 A_n]| \Psi_n \rangle.
    \end{align}
While this quantity theoretically approaches ${}^2 D_n (\epsilon_{n-1})$ as $\epsilon_n \rightarrow 0^+$, in practice the discrete unitary gates are subject to substantial noise on current-to-intermediate-term quantum computers and hence, do not collapse to the identity operator for any actual gate sequence. The noise channels in general will contract the set of possible $2$-RDMs (e.g., for systems with strong depolarizing errors this is to a fully depolarized 2-RDM). For our system, this can lead to a result that any energy obtained will be higher than the energy of the previous step (see Supplemental Material for an example). In these instances, the change in energy due to noise is greater than any change from the optimization.

Let ${}^2 \Gamma_n$ be a matrix of the same rank as the 2-RDM, and ${}^2 \tilde{D}_n(\epsilon_n)$ be the corrected 2-RDM. Then we define a correction by the following system of equations:
\begin{align}
    \label{em1} {}^2 \tilde{D}_{n+1} (\epsilon_{n}) &= {}^2 D_{n+1}(\epsilon_{n}) + \sum_{i=0}^{n} {}^2 \Gamma_ i   \\
    \label{em2} {}^2 \Gamma_{n} &=  {}^2 D_{n}(\epsilon_{n-1}) - {}^2 D_{n+1}(0^+) ,  \\
    \label{em3}{}^2 D_0 &= {}^2 D_{\rm HF}.
\end{align}
Eq.~\eqref{em1} defines the error mitigated 2-RDM at each step.  The ${}^2 \Gamma_n$ in Eq.~\eqref{em2} is the difference between the new state with $\epsilon_n=0^+$ and the previous state.  Eq.~\eqref{em3} gives the initial condition of the system. The correction helps to avoid \tc{noise-related barriers} in the optimization surface (as ${}^2 \tilde{D}_{n+1}(0^+)={}^2 D_n(\epsilon_{n-1})$), allowing us to reach 2-RDMs that are normally inaccessible due to the noise. For a noise-free simulation, we also have that ${}^2 \Gamma_n=0$ for all $n$, ensuring that we would maintain the exact result on a perfect quantum computer.  We use the corrected 2-RDM ${}^2 \tilde{D}_{n+1}$ throughout the optimization in evaluating the energy as well as choosing the elements of ${}^2 A_{n+1}$. While the gradient information \tc{reflected in ${}^2 A$} around ${}^2D$ and ${}^2 \tilde{D}$ will not be the same when ${}^2 \Gamma$ is large, because we are optimizing $E[{}^2 \tilde{D}_n]$, and because we generate ${}^2 A_n$ with Eq.~\eqref{classA}, this is the appropriate choice. If we were to use Eq.~\eqref{quantA} instead, then we would obtain information around ${}^2 D_n$, and would have to correct ${}^2 A$ as well.

There are a number of practical considerations in the implementation of the ${}^2 \Gamma$-correction such as the potential variability of the noise. Because we are adding RDMs with separate uncertainties, the uncertainty in the result increases (if we assumed independent ${}^2\Gamma_i$ with equal standard deviations $\sigma$, this would be $\sqrt{n}\sigma$ after $n$ iterations), which may require us to increase the sampling of ${}^2 \Gamma_i$. The errors affecting the quantum computer may exhibit a time dependence on the order of the run time. To avoid this possibility, we run the results as contiguously as possible with the total number of iterations $n$ being kept relatively low (for all instances $n \le 5$). Additionally, the 2-RDM is purified in some cases to ensure that the negative eigenvalues of the 2-RDM and the related 2-hole and particle-hole RDMs (see next section) are eliminated. Regardless, we find this error mitigation strategy to be necessary to obtain meaningful results within the context of an iterative ansatz.

\subsection{Purification of the 2-RDM}

As mentioned above, the effect of noise in a quantum simulation is that measured quantum state might no longer represent a physical system. While we cannot directly assess the purity or fidelity of an RDM, we can ``purify'' the 2-RDM to ensure that the eigenvalues of the various permutations of the particle and hole reduced density matrices are positive semidefinite, which are necessary criteria for a pure-state or ensemble $N$-representable 2-RDM~\cite{Mazziotti2002pre}. A matrix is positive semidefinite if and only if its eigenvalues are nonnegative.  For instance, for the 2-RDM, the 2-particle (${}^2 D$), 2-hole (${}^2 Q$), and particle-hole (${}^2 G$) matrices must have nonnegative probabilities, and hence, must be positive semidefinite in a set of conditions known as the 2-positivity (or DQG) conditions~\cite{Garrod1964, Coleman1963, Erdahl1978, Foley2012}
\begin{align}
    {}^2 D & \succcurlyeq  0, \\
    {}^2 Q & \succcurlyeq  0, \\
    {}^2 G & \succcurlyeq  0,
\end{align}
where the elements of these metric matrices are given by
\begin{align}
    {}^2 D^{ij}_{kl} &= \langle \Psi |{\hat a}^\dagger_i {\hat a}^\dagger_j {\hat a}^{}_l {\hat a}^{}_k | \Psi \rangle, \\
    {}^2 Q^{kl}_{ij} &= \langle \Psi |{\hat a}^{}_k {\hat a}^{}_l {\hat a}^{\dagger}_j {\hat a}^{\dagger}_i | \Psi \rangle, \\
    {}^2 G^{il}_{kj} &= \langle \Psi |{\hat a}^\dagger_i {\hat a}^{}_l {\hat a}^{\dagger}_j {\hat a}^{}_k | \Psi \rangle .
\end{align}
We accomplish the purification by semidefinite programming, which allows us to minimize an function of a matrix subject to linear constraints while ensuring that the matrix remains positive semidefinite \cite{VANDENBERGHEt1996, Mazziotti2011,Mazziotti2004_sdp_qc}. The general method was developed by one of the authors for reconstructing noisy processes for quantum tomography~\cite{Foley2012}, \tc{and was more recently applied in the context of quantum simulation by Rubin et al. \cite{Rubin2018}.}

The objective in this work is to create a purified 2-RDM, ${}^2 D_{\rm SDP}$, which minimizes the norm of the error matrix $E = {}^2 D - {}^2 D_{\rm SDP}$, subject to the constraints (DQQ) ensuring that ${}^2D_{\rm SDP}$ represents a physical system. To express this as a semidefinite program, we take $F$ to be a matrix of free variables, and then minimize the trace of the following block matrix:
\begin{equation}
    \begin{pmatrix}
    I & E \\ E^{\dagger} & F
    \end{pmatrix}
    \succcurlyeq  0 .
\end{equation}
Taking the determinant of the $2 \times 2$ block matrix allows us to relate the trace of $F$ to the Frobenius norm, providing a semidefinite relaxation for the minimization problem. The DQG constraints can be expressed in a block-diagonal form:
\begin{equation}
    \begin{pmatrix}
    {}^2 D &0 &0 \\ 0 & {}^2 Q & 0 \\ 0 & 0 & {}^2 G
    \end{pmatrix} \succcurlyeq 0 .
\end{equation}
These semidefinite conditions, the linear mappings between the metric matrices, and the trace of the 2-RDM define the constraints in the SDP. To solve the SDP, we use a boundary-point algorithm, developed by one of the authors for the direct variational calculation of the 2-RDM~\cite{Mazziotti2004_sdp_qc, Mazziotti2007_book, Montgomery2018, Boyn2020}.  The algorithm for purification of the 2-RDM with the DQG conditions has a scaling of $O(r{}^6)$.

\section{Benzyne Calculations}

In this work we use the qACSE method to investigate the ortho-, meta-, and para- isomers of benzyne, which may be obtained via the elimination of two substituents in the relevant positions of the benzene ring. Owing to their versatility as reactive intermediates in biological processes, derivatives of the isomeric benzynes have been the subject of a growing interest in the synthetic research community in the development of biomimetic reactions~\cite{Wolfram1999}, such as the Diels-Alder reaction~\cite{Hoye2017} and in so-called ``click chemistry''~\cite{Larock2008}, with a wide range of applications to the synthesis of heterocycles~\cite{Larock2013} and natural products~\cite{Hoye2017}. Even though biradicals such as benzyne play key roles across synthetic and materials chemistry, making their accurate theoretical description quintessential to the understanding of chemical processes, their exact treatment continues to pose a challenge to electronic structure theory~\cite{Shee2019, Yang2015}. \tc{Details regarding the electronic structure treatment of these systems are included in the Appendix. It also is important to note that one of the symmetries used in the 5-qubit reduction is approximate for the meta-isomer, and leads to an error of approximately 0.008 hartrees versus the FCI result. }

Figure~3 shows the structures for each of the three isomers, as well as the occupations of the highest and lowest occupied natural orbitals. The energetic ordering of the three isomers follows their degree of diradical character, with experimental gas phase heats of formation showing ortho as the energetically lowest isomer, followed by the meta and then para isomers, at energies of  $10 \pm 3$ kcal/mol and $22 \pm 3$ kcal/mol relative to the ortho reference, respectively~\cite{Lineberger1998}. The variations in ground-state energy and diradical character are driven by the degree to which the geometric constraints of the given isomer allow for overlap between the singly occupied carbon-p orbitals, which is demonstrated by the electron densities of the highest occupied natural orbital (HONO) and the lowest unoccupied natural orbital (LUNO), shown in Fig.~3.  In the ortho isomer, adjacency of the singly occupied orbitals allows for good overlap and energetically favorable formation of a bond with significant $\pi$ character, giving this isomer C-C triple bond character. While somewhat compensated by geometric distortion, driven by the greater C-C radical distance the magnitude of this bonding interaction is reduced in the meta isomer, and essentially diminished in the para geometry, where no overlap between the lobes of the carbon-based radical orbitals is geometrically feasible.

\begin{figure}[h]
    \centering
    \includegraphics[scale=0.20]{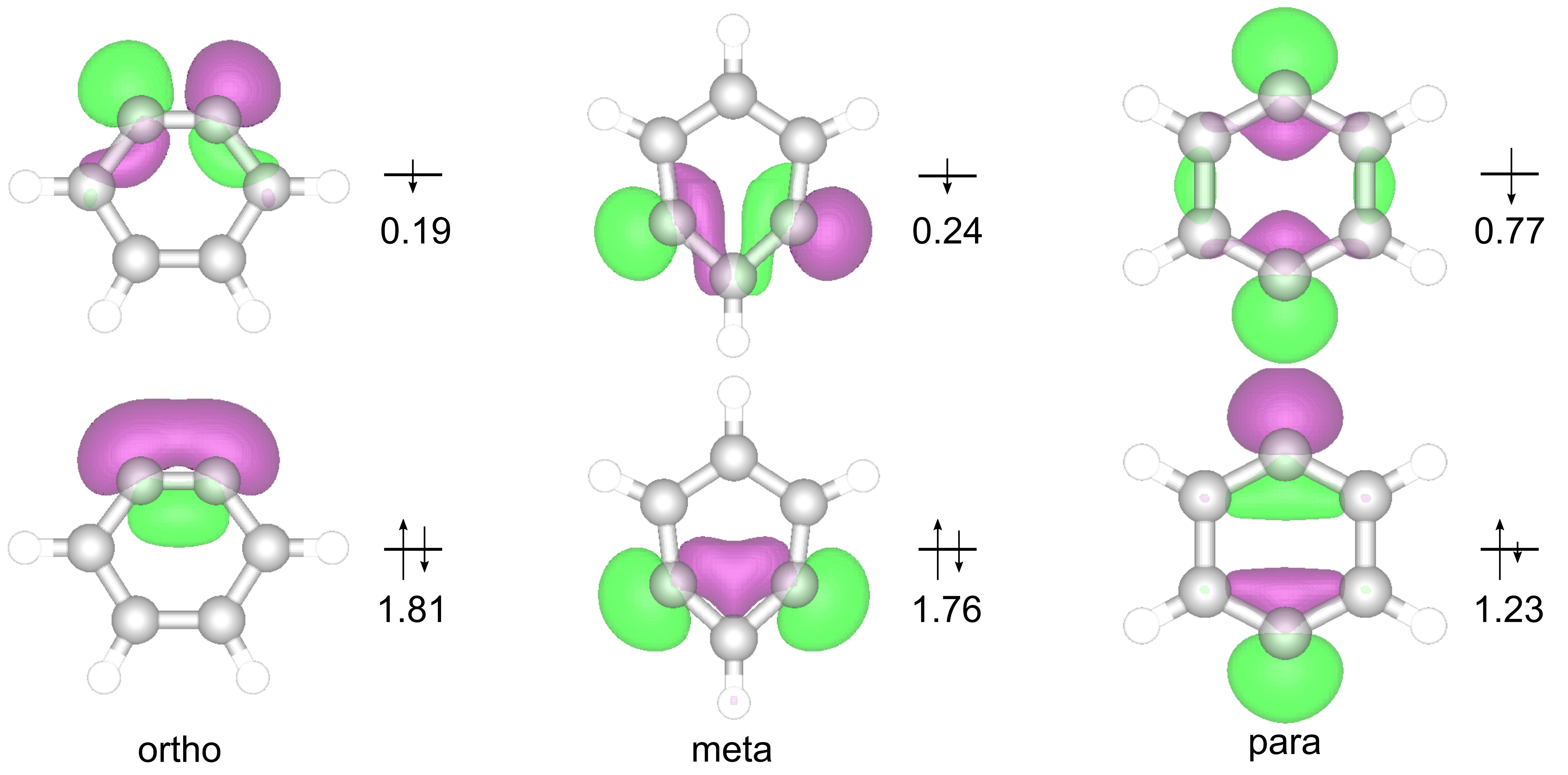}
    \caption{Molecular orbital diagram and natural-orbital occupations of the highest and occupied lowest natural orbitals for ortho-, meta-, and para-benzyne. Geometries for the ortho- and meta- isomers were obtained from reference~\cite{Krylov2012} and optimized with \tc{spin-flip time dependent density functional theory (SF-TDDFT)}, and the para- isomer was obtained from reference~\cite{Krylov2002} and optimized with \tc{spin-flip coupled cluster with singles and doubles (SF-CCSD).}}
    \label{fig:benzyne_mos}
\end{figure}
\begin{figure}[h]
    \centering
    \includegraphics[scale=0.75]{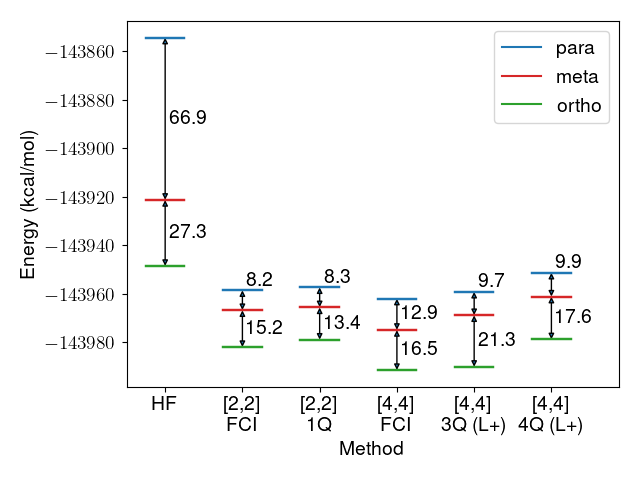}
    \caption{Overview of results shows active space calculations for the different configurations of benzyne across several methods, including Hartree-Fock, and full configuration interaction (FCI) and qACSE for [2,2] and [4,4] active spaces. The 3- and 4-qubit results utilize the limit-preserving correction (L) and purification (+) schemes of error mitigation. The data corresponds with results taken in Table~I.}
    \label{fig:rel_ens}
\end{figure}

The relative energies from the complete active space self-consistent field method (CASSCF) and from the quantum calculations are listed in Table~I and Figure 4 for the [2,2] and [4,4] active spaces where the notation [X,Y] denotes X electrons in Y orbitals.  The CASSCF calculations optimize the active electrons and orbitals in the mean-field of the remaining electrons and orbitals.  The target CASSCF results yield the correct ordering, although each gap is slightly higher than experimental values. For the [2,2] case, the meta and para energies relative to ortho are 15 and 23~kcal/mol, respectively. For the [4,4] active space, the meta and para energies relative to the ortho configuration are 13 and 29~kcal/mol respectively.  The [2,2] active space corresponds with a 1-qubit quantum calculation, whereas the [4,4] calculation was performed with 3 and 4 qubits. The error mitigation ranges from a simple measurement correction to our full scheme of corrections (denoted \tc{L+, or MPL+}). $M$ refers to a state preparation and measurement, $P$ to the application of the number projection, $L$ to the use of the ${}^2 \Gamma$-correction, and $+$ to the SDP corrected state. The error in the obtained relative energies on the quantum computer in the [4,4] case is less than 5 kcal/mol (8 mhartree) for both the 3-qubit (3Q) and 4-qubit (4Q) cases, whereas for the [2,2] space, we obtain a result within 2~kcal/mol (3~mhartree). The number of unique iterations is between 2$-$5, depending on the ansatz developed. The \tc{operators} (see Supplemental Material) for the 3-qubit calculations have 2$-$4 CNOT gates, while the pool of operators for the 4-qubit operators each have 8-12 CNOT gates.

\begin{table}[h]
    \caption{Relative energies between the configurations of benzyne with CI (CASCI) and qACSE methods for differing active spaces and levels of error mitigation, in kcal/mol. [0,0] active space refers to the initial Hartree-Fock calculation. $M$ refers to a state preparation and measurement error, P to the application of the number projection, L to the use of the ${}^2 \Gamma_n$-correction, and $+$ to the SDP corrected state.}
    \label{relative}
    \centering
    \begin{tabular}{c|>{\centering\arraybackslash}  m{2cm}  >{\centering\arraybackslash}  m{2cm}}
\hline
 \multirow{2}{*}{[AS], $N_{Q}$, Err. Mit.}  & \multicolumn{2}{c}{$E_{\rm conf} - E_{\rm ortho}$ (kcal/mol)}  \\
&  \em{meta} & \em{para} \\
\hline
\multicolumn{3}{c}{Configuration Interaction}  \\
\hline
\relax [0,0] & 27.3 & 94.2 \\
\relax [2,2] & 15.2 & 23.4  \\
\relax [4,4] & 16.5 & 29.5  \\
\hline
\multicolumn{3}{c}{qACSE} \\
\hline
\relax [2,2], 1, M  &  13.4 & 21.7 \\
\relax [4,4], 3, MP &  3.2 & 15.4 \\
\relax [4,4], 3, MPL&  47.8 & 55.7 \\
\relax [4,4], 3, MPL+& 21.3 & 31.0  \\
 \relax [4,4], 4, MPL & 27.1 & 23.6  \\
\relax [4,4], 4, MPL+ & 17.6 & 27.5 \\
\hline
Experiment~\cite{Lineberger1998} &  10 $\pm$ 3 & 22 $\pm$ 3   \\
\hline
    \end{tabular}

\end{table}
Another comparison between the error mitigation schemes is seen in the target energies for each calculation relative to the CI result. These errors are listed in Table II. In particular, despite having differences between configurations of only a few kcal/mol, the difference from the FCI results for results without the ${}^2 \Gamma$-correction is around 20 kcal/mol higher than the target energies across the configurations. These results are more common for what might be expected from noisy quantum devices, as often the lowest energy states are not the final state of the optimization. The ${}^2 \Gamma$-corrected results on the other hand in some instances can be below the variational CI bound, highlighting the need for purification.

\begin{table}[h]
    \caption{Difference in energy between the qACSE methods including various error mitigation schemes and the FCI result in millihartrees (mhartree).}
    \label{tab:absolute}
    \centering
    \begin{tabular}{c|>{\centering\arraybackslash}  m{1.5cm}>{\centering\arraybackslash}  m{1.5cm}>{\centering\arraybackslash}  m{1.5cm}}
\hline
\multirow{2}{*}{[AS], $N_{Q}$, Err. Mit.} & \multicolumn{3}{c}{Error versus CI (mhartree)} \\
  &  \em{ortho} & \em{meta} & \em{para}  \\
\hline
\relax [2,2], 1, M  & 4.8 & 1.9 & 1.9   \\
\relax [4,4], 3, MP & 51.9 & 30.6 & 29.5  \\
\relax [4,4], 3, MPL& -53.6 & -3.8 & -11.8  \\
\relax [4,4], 3, MPL+& 2.1 & 9.7 & 4.5 \\
\relax [4,4], 4, MPL & -25.9 &  -9.0 & -35.3  \\
\relax [4,4], 4, MPL+ & 20.2 & 21.9 & 17.0\\
\hline
    \end{tabular}
\end{table}

Finally, the natural-orbital occupation numbers, which are the eigenvalues of the 1-RDM, can help infer the nature and degree of electron correlation in the system. The Hartree-Fock state, corresponding with a single determinant, has eigenvalues of $2$ or $0$ across all (spatial) orbitals, while a biradical system would exhibit equal occupations of 1 in the highest occupied and lowest unoccupied natural orbitals. We report the natural orbital occupations for the FCI and purified results in Table III for the $1$-,$3$-, and $4$-qubit qACSE calculations. In each case, we see significant differences between the para isomer and the other two isomers (ortho and meta) on the quantum computer. The para-benzene, which does not have any overlapping density between the carbon $p$ orbitals (see Fig.~3), exhibits biradical character, whereas the other two configurations exhibit more single-reference character. This is also reflected in the amount of correlation energy recovered ($E_{\rm FCI}-E_{\rm HF}$) for each configuration (see Fig.~4). When compared to the FCI occupations, the results for the 3-qubit case were all with 0.09 of the target occupations. In the 4-qubit case the ortho ($0.14$) and meta ($0.19$) HONO and LUNO occupations have more significant errors, which could be expected from the increased absolute energies seen for each of these isomers. By looking at the HONO-1 and LUNO+1 orbitals in the [4,4] space, we also see that the fractional occupations of the HONO and LUNO are not an artifact of error on the quantum computer, as the closeness of the HONO-1 and LUNO+1 occupations to 2 and 0 is maintained.

\setlength{\tabcolsep}{5pt}
\begin{table}[h]
        \caption{Largest natural-orbital occupation numbers for the FCI results and the purified, ${}^2 \Gamma$-corrected results for the [2,2] and [4,4] active spaces on the quantum computer. In each case, the para-benzyne solution exhibits biradical character in the highest occupied and lowest unoccupied natural orbitals, though to differing degrees based on the method.}
    \label{tab:NOON}
    \centering
    \begin{tabular}{cc|ccc}
    \hline\multirow{2}{*}{Method ($N_Q$)}&  \multirow{2}{*}{Orbital} & \multicolumn{3}{c}{Orbital Occupations}
    \\
    & & \em{ortho} & \em{meta} & \em{para} \\    \hline
    \relax \bf{FCI [2,2]}
    & HONO & \rm{1.811} & \rm{1.712} & \rm{1.232}  \\
    & LUNO & \rm{0.189} & \rm{0.288} & \rm{0.768} \\
    qACSE (1)
    & HONO & 1.695 &	1.604	& 1.127  \\
    & LUNO & 0.305	& 0.396& 0.873 \\
    \hline \relax
    \bf{FCI [4,4]}
        & HONO$-1$& \rm{1.947} & \rm{1.977} & \rm{1.981} \\
        & HONO    & \rm{1.813} & \rm{1.756} & \rm{1.235} \\
        & LUNO    & \rm{0.187} & \rm{0.244} & \rm{0.765} \\
        & LUNO$+1$& \rm{0.053} & \rm{0.023} & \rm{0.019} \\
    \relax qACSE (3)
        & HONO$-1$ & 1.956	&1.994&	1.992 \\
        & HONO & 1.851	&1.716 &	1.148  \\
        & LUNO & 0.149	&0.283	&0.852 \\
        & LUNO$+1$ & 0.045	&0.006	&0.008 \\
    \relax qACSE (4)
        & HONO$-1$ & 1.973	&1.985 	&1.956  \\
        & HONO & 1.938	& 1.570& 1.200 \\
        & LUNO & 0.054	& 0.433 &	0.790 \\
        & LUNO$+1$ & 0.036 &	0.012 & 0.055 \\
    \hline
    \end{tabular}

\end{table}

\section{Discussion}
The results of these benzyne calculations highlight the potential for quantum simulation on near-term devices, particularly with quantum RDM methods and error mitigation tools designed for RDMs. Though work in our group and elsewhere has investigated and obtained highly accurate results for small systems or particular configurations of electrons (namely in taking advantage of pure $N$-representability constraints)~\cite{Smart2019hqca,Smart2020geminal,Arute2020}, this work represents a step towards more general quantum computing algorithms based on RDM theory. Indeed, the [4,4] active space represents an important step from model systems and minimal cases towards the end goal of robustly treating strongly correlated many-body systems. These results also demonstrate a useful classical-quantum hybrid approach, incorporating elements from both classical and quantum techniques.

\tc{With regards to the number of iterations and the variational cost, for many systems, including the benzyne isomers,} the qACSE \tc{method} is consistently able to recover a large part of the correlation energy within a few iterations. \tc{However, there are systems that have search directions in shallow gradients where ACSE algorithm may have difficulties.} While we also evaluated derivative-free 1-dimensional optimizers that might be able to help in a noisy landscape \cite{Peruzzo2014,Smart2019hqca,Robinson2006}, practically, the trust-region optimization combined with \tc{a} rejection criteria provides a reliable way of \tc{choosing a step size for Eq. (4), and making sure that convergence progresses as a whole}. The rejection criteria in particular eliminate iterations which do not contribute to the ansatz properly with an optional reevaluation of the last ${}^2 A_n$ step. This helps in particular with overcoming instances where the errors in the gradient are too large to take a meaningful step. \tc{It is also worth mentioning that the experiment requirements for convergence and termination of the method are different from ideal conditions. While lowering the residuals of the ACSE is ideal, and ensures a properly converged state, noise will combat the ability of actually reaching a meaningful RDM. } \tc{Because of the limitations of noise, in the present multi-qubit examples the ${}^2 A $ matrix is updated by a classical algorithm with reconstruction of the 3-RDM rather than the quantum algorithm shown in Eq.(\ref{quantA}). In these instances, error from reconstruction of the 3-RDM is lower than the error from the noise on the quantum devices.  Importantly, the classical and quantum algorithms can be interchanged depending upon the complexity of the circuit and the level of noise on a given device.}

\tc{These results highlight the necessity of different error mitigation, which we discuss in a somewhat qualitative manner based on hand-on experience with the benzyne system. In other systems, some of these techniques might behave differently or be more or less impactful. The qubit reduction technique (see Appendix B. 3) allows for significant simplification of the problem (although not to a trivial degree for the [4,4] case), as well as for different thresholds of accuracy. We restate something that is somewhat known, that with the Jordan-Wigner transformation and $r$ spatial orbitals, one can always find two $Z$ symmetries of length $2r$ and $r$, corresponding to the parity of the $\alpha$ and $\alpha+\beta$ sets of orbitals, which reduces the number of qubits to $2r-2$. We did not explicitly identify the effect of the measurement errors involving the inversion of the state transition matrix, although these have been documented elsewhere to help improve results on the order of the measurement error (0.01). Because incorrectly measured states can easily lead to different particle states, this can lead to large differences in the obtained energies. However, regardless of the measurement error, the projection of the RDM onto the correct particle number space in the diagonal entries is a critical step. The energetic effect of this correction is system dependent, but can easily be on the order of hartrees. Quite simply put, the results are often not meaningful without this correction, which can also be seen in its success in other work\cite{Smart2020geminal, Arute2020}. While it is preferable in theory to correct the diagonal and off-diagonal elements of the 2-RDM, for the latter instances, a measurement sequence which commutes with the particle number operator must be developed. Additionally, this greatly changes the tomography requirements of the 2-RDM, rendering useless the advantages of local measurement commutation. The incremental improvements in the quantum devices over the last year are also critically important, as other devices were tested that did not achieve the same level of results (not reported).}

The ${}^2 \Gamma$-correction serves indirectly to expand the set of accessible 2-RDMs while preserving the integrity of the iterative optimization. While the application here to an iterative ansatz is unique, the idea at each iteration could be seen as a zeroth-order extrapolative procedure, like the Richardson extrapolation, repeated at each iteration~\cite{Temme2017, Kandala2018}. Instead of attempting a linear or higher order fit to a variable noise strength, we simply add a correction RDM. As a result, we do not have to deal with adjusting how noise is applied in the underlying pulse, and the cost of the mitigation procedure is kept low. Even if at each step we recalculated ${}^2\Gamma$, the number of evaluations would be linear with respect to $n$. While the implementation here is straightforward, it is likely that this method or variations on it could be applied to other iterative methods in a straightforward manner. In terms of the set of possible RDMs that can be measured, this approach slowly shifts our corrected RDM by ${}^2 \Gamma$ through the set of all possible RDMs. \tc{Qualitatively, the effect of this strategy on the obtained energy is to improve the result usually by tens of mhartrees, and in some instances up to 0.1 Hartree. However,} as it is possible to move beyond the boundary of the set of physical RDMs, purification of the RDM is a necessary step, \tc{albeit with approximate $N$-representability conditions. The distance between the ${}^2 \Gamma-$corrected 2-RDM and the purified 2-RDM, which is also not consistent, can be used as an exclusionary criteria in the optimization.}

\tc{Both qACSE and ADAPT-VQE use the ACSE wave function ansatz~\cite{Mazziotti2007_pra, Mazziotti2004, Mazziotti2020, Mazziotti2006_prl}  that was developed in the ACSE literature~\cite{Mazziotti2006_prl, Mazziotti2007_pra, Gidofalvi2009, Mukherjee2001} (for example, see section IIE of Ref.~\onlinecite{Mazziotti2007_pra}).  The structure of this wave function---product of unitary two-body exponential operators on a reference wave function---has the ACSE as its stationary equation~\cite{Mazziotti2007_pra, Mazziotti2004}.   The ACSE ansatz is related to the single-term two-body exponential ansatzes~\cite{Nakatsuji2000, Nooijen2000, Voorhis2001, Nakatsuji2001, Davidson2003, Ronen2003, Piecuch2003, Kutzelnigg2005} and the two-body exponential product ansatzes~\cite{Mazziotti2004, Mazziotti2020}, which were investigated in the context of the contracted Schr{\"o}dinger equation (CSE)~\cite{Mazziotti1998b, Nakatsuji1996, Yasuda1997, Colmenero1993a, Valdemoro2008, Mazziotti2002d, Mazziotti1999a, Coleman2000}.  Notably,  while this wave function has been stated heuristically and called an adaptive generalized unitary coupled-cluster singles and doubles wave function in the ADAPT-VQE literature, its stationary equation is not a coupled cluster equation, and its definition in the ACSE literature significantly predates its recent discussion.  In fact, Grimsley {\em et al.}~\cite{Grimsley2018} describe ADAPT-VQE as ``not so much an approximation to UCC [unitary coupled cluster] as it is a wholly unique ansatz.’’  From this perspective, by minimizing the ACSE wave function, both qACSE and ADAPT-VQE are seeking solutions of the ACSE---rather than a direct solution of the Schr{\"o}dinger equation as in VQE, and hence, both can be understood as types of contracted quantum eigensolvers.  The distinction between the VQE and CQE is important because the CQE framework informs both the structure of the wave function and its stationarity condition.}

\tc{Although both qACSE and ADAPT-VQE can be viewed as quantum solutions of the ACSE, their initial implementations have some significant differences.  The ADAPT-VQE~\cite{Grimsley2018} defines a predefined pool of parameterized unitary two-body exponential operators from which the ACSE wave function can potentially be constructed from the reference (Hartree-Fock) wave function. The algorithm improves the trial ACSE wave function at the $n^{\rm th}$ iteration by ({\em i}) multiplying the $(n-1)^{\rm th}$ ACSE wave function by the operator from the pool with the largest energy gradient and ({\em ii}) reoptimizing the energy with respect to all parameters in the pool operators.  In contrast, the qACSE does not use a predefined pool of operators but rather computes the residual of the ACSE either from a an efficient quantum measurement of an effective 2-RDM as shown in Eq.~(10) or a classical evaluation where the 3-RDM is approximately reconstructed.  This generality gives the qACSE additional flexibility, which may become increasingly important in the treatment of larger, more correlated atoms and molecules where a limited operator pool may miss significant correlation effects.  Moreover, the qACSE does not reoptimize its parameters in previous steps as in part ({\em ii}) of the ADAPT-VQE algorithm.  While a reoptimization phase decrease circuit depth, especially for small molecules, it is not necessary for converging to a solution of the ACSE, and it may require a significantly larger number of energy function and gradient evaluations for larger molecules.}

\section{Conclusions}

Molecular simulations on quantum computers have the potential to treat strongly correlated problems that are currently intractable on conventional computers.  The practical realization of such simulations, however, requires quantum molecular algorithms that are mappable to transformations, such as products of unitary transformations, that are natural for quantum computers.  \tc{Here we implement a novel solution from a contraction of the Schr{\"o}dinger equation onto the space of only two electrons, known as the anti-Hermitian contracted Schr{\"o}dinger equation (ACSE).} To make the solution of the ACSE more practical for more realistic chemical problems on quantum computers, we \tc{utilize robust error mitigation techniques, including techniques} based on $N$-representability constraints.  The solution of the anti-Hermitian CSE (ACSE) through iterative minimization of its residual generates a rapidly convergent product of two-body unitary transformations that is natural for implementation on quantum computers.  Furthermore, unlike the solution of the ACSE on the classical computer, the contracted Schr{\"o}dinger solver on quantum computers \tc{can fully or partially remove approximate reconstructions of} higher RDMs, and hence, can potentially achieve exact results without the exponential complexity of the many-electron wave function.

The combination of the ACSE solver with \tc{robust error mitigation} provides a scalable approach to molecular simulations on quantum computers with low circuit depth and few variational parameters.  We apply the algorithm to the resolution of  the {\em ortho-}, {\em meta-}, and {\em para-}isomers of benzyne ${\textrm C_6} {\textrm H_4}$.  The relative energies exhibit single-digit millihartree errors, and the computed natural-orbital occupations capture the biradical nature of the {\em para}-isomer.  The molecular simulation of the benzyne isomers represents an important step in eigensolver and error-mitigation technologies towards the practical simulation of larger, even more complex molecules on quantum computers.

\begin{acknowledgments}
D.A.M. gratefully acknowledges the Department of Energy, Office of Basic Energy Sciences, Grant DE-SC0019215 and the U.S. National Science Foundation Grants No. CHE-2035876, No. DMR-2037783, and  No. CHE-1565638. The views expressed are of the authors and do not reflect the official policy or position of IBM or the IBM Q team. \tc{We also are grateful for the reviewers in providing helpful suggestions that improved the manuscript.}
\end{acknowledgments}

\appendix
\section{Electronic Structure Calculation}
Complete active state self consistent field (CASSCF) calculations were performed as implemented in the Maple Quantum Chemistry Package~\cite{Maple, QCP, PySCF} using [2,2] and [4,4] active spaces with the correlation-consistent valence double-zeta (cc-pVDZ) basis set~\cite{ccbasis}. Following convergence of the CASSCF procedure, effective active space electron integrals for the quantum ACSE calculation were obtained via the folding of the core-active cross terms into the active space, such that effective active space energy is given by:
\begin{equation}
\Tilde{E}_{\text{act}} = \frac{1}{2} \sum_{pqst} {}^2\Tilde{K}^{pq}_{st} \, {}^2D^{pq}_{st}   \,,
\end{equation}
where ${}^2\Tilde{K}^{pq}_{st}$ are the active space electron integrals containing the core-active cross terms. The elements of the effective active space integral matrix ${}^2\Tilde{K}^{pq}_{st}$ are constructed from the one- and two-electron integrals as follows:
\begin{equation}
    {}^2\Tilde{K}^{pq}_{st} = \frac{1}{N-1}(\,{}^1\Tilde{K}^p_s\delta^q_t + \,{}^1\Tilde{K}^q_t\delta^p_s) + {}^2K^{pq}_{st} \,,
\end{equation}
where
\begin{equation}
    {}^1 \Tilde{K}^p_s = {}^1 K^p_s + \sum_i (2 \,{}^2K^{pi}_{si} - {}^2 K^{pi}_{is}) \,,
\end{equation}
and $p,q,s,t$ runs over all active orbitals and $i$ runs over all core orbitals.

\section{Quantum Calculation}
Using the electron integrals for the active space from above, we perform a quantum calculation on different IBMQ devices. In particular, we perform [2,2] and [4,4] calculation under the Jordan-Wigner transformation. Different IBMQ devices were utilized through the IBM Quantum Experience. These devices utilize fixed-frequency transmon qubits with co-planer waveguide resonators~\cite{Koch2007,Chow2011}. We use the \textsc{python} 3 package \textsc{qiskit} (v 0.15.0)~\cite{Qiskit} to interface with the device. The calculations themselves are multifaceted, with nonstandard approaches taken in a number of different areas. We document these in subsequent sections. Each measurement was performed with $2^{13}$ shots. Stochastic effects were on the order of mhartree, \tc{though these are somewhat of a lesser concern due to the purification scheme}. For the collection of all 2-RDMs we utilized a symmetry projected operator basis using the $\hat{N}$ and $\hat{S}_z$ symmetries~\cite{Smart2021tomo}.

For the 1-qubit calculations, we utilized ibmq-armonk, while for the 3- and 4-qubit calculations, we utilized ibmq-bogota, a 5-qubit superconducting device. \textsc{qiskit} was used to interface with the IBMQ devices.

\subsection{Quantum $[2,2]$ Active Space Calculations}
Using the Jordan-Wigner transformation, the [2,2] case with 4 spin orbitals maps to 4 qubits. The [2,2] calculations contain two Pauli symmetries related to the parities of the total number of electrons and the number of electrons in a subset of spin orbitals (either $\alpha$ or $\beta$), and a further symmetry is found for most molecular systems, allowing the [2,2] system to be represented with a single-qubit. These can be expressed as:
\begin{align}
S_1 = \{ Z_1 Z_2, Z_1 Z_3 , Z_1 Z_4  \}.
\end{align}
The elements of ${}^2 A$ were determined through the quantum ACSE method, with Euler's method being used to propagate the ansatz. An $l_{2}$ norm of ${}^2 A$ \tc{below 0.01} was used as the stopping criterion, which was usually reached in 10-12 iterations. The exact exponential of any combination of Pauli operators is well known for the single-qubit case, and so we are able to exactly express $U = \prod_i e^{A_i}$ as well as $U' = e^{iH\delta }\prod_i  e^{A_i}$. For these runs, we chose $\delta=0.25$.

\subsection{Quantum $[4,4]$ Active Space Calculations}
The Jordan-Wigner representation maps the [4,4] case with 8 spin orbitals to 8 qubits. Again, two symmetries related to fermionic parity can be utilized, and then depending on the Hamiltonian we can find additional symmetries. For these particular integrals, we find 2 additional symmetries across all configurations, and then an additional symmetry for the para- configurations, which describes the ortho- and meta- configurations at integral cutoff threshold of between $1.0-1.1  \times 10^{-3}$ hartrees and $3.1-3.2 \times 10^{-3}$ hartrees, respectively. The difference from the target (FCI) energy in the \tc{3-qubit} ortho- case is less than $9.2 \times10^{-4}$ H, and in the \tc{3-qubit} meta case, approximately $8.6 \times 10^{-3}$ hartrees.

The symmetries are listed in the follow set for the 4- and 5-cases respectively:
\begin{align}
    S_4 = \{  Z_1 Z_2Z_3 Z_4, Z_1 Z_2 Z_5 Z_6, Z_1 Z_3 Z_5 Z_6, Z_2 Z_3 Z_5 Z_8 \}
\end{align}
\begin{align}
    S_3 = \{Z_1 Z_2 Z_3 Z_4, Z_1 Z_5, Z_2 Z_6, Z_3 Z_7, Z_1 Z_2 Z_3 Z_8 \}.
\end{align}
As a result, we are able to perform 3- and 4- qubit simulations of these systems on the 5-qubit linearly connected ibmq-bogota device. When tapering off qubits, we use eigenvalues which match the eigenvalues of the initial closed-shell singlet Hartree-Fock determinant.

The calculations themselves utilized a model-trust region Newton's method, where the initial trust region was taken to be 2, and the quadratic fit was taken from $\epsilon_n = \pm 1$. Additionally, we used a threshold of $0.75\times a_{max}$ where $a_{max}$ indicated the largest magnitude term in the ${}^2 A$ for a given iteration. The convergence criteria was taken to be 0.02 in the trust region criteria, and we used 5 iterations as the maximum allowed iterations, which generally yielded 2-4 terms in the ansatz.

\subsection{Qubit Reduction by Tapering}
The qubit reduction scheme follows previous work by Bravyi et al. and expanded by Setia et al.~\cite{Bravyi2017, Setia2019}. In particular, we express the Hamiltonian in the Pauli basis and then put these terms in a check sum representation to construct the generator and parity check matrices from the field of quantum error correction~\cite{Gottesman1997}. By performing Gaussian elimination on the parity check matrix, we can find generators of the Hamiltonian, which in turn allow us to select a basis for the corresponding null space. Elements of the null space will commute with every term in the Hamiltonian, and thus are symmetries of $H$. By using a unitary transformation:
\begin{equation}
    U_i = \frac{1}{\sqrt{2}}(X_j+s_i)
\end{equation}
where $X_j$ is selected so that $X_j$ anticommutes with $s_i$, and commutes with all other $S_i,~ i \neq j$,  we transform the Hamiltonian so that qubits $j$ have only $X$ or $I$ in each term. By selecting an appropriate eigenvalue of $X$, we can taper off these terms, resulting in a modified fermionic transformation. We use eigenvalues which agree with the eigenvalues of the initial closed-shell singlet Hartree-Fock determinant.

\subsection{Classical Solution to the ACSE}
In the fully quantum algorithm, the quantum computer is used in both the calculation of the ${}^2 A $ and ${}^2 D$ matrices. For the [4,4] cases we used a classical approach in solving for elements of ${}^2 A$, which reduces the computational demands on the quantum computer and yields sufficient accuracy in this case. This can be found by calculating elements of ${}^2 A$ from:
\begin{equation}
    {}^2 {A}^{i,k}_{j,l} = \langle \Psi | [{\hat a}^\dagger_i {\hat a}^\dagger_k {\hat a}^{}_l {\hat a}^{}_j,\hat{H}]| \Psi \rangle.
\end{equation}
More specifically, for a molecular system, the reduced Hamiltonian ${}^2K$ can be written as:
\begin{equation}
{}^2 K^{p,r}_{q,s} = \frac{1}{2(N-1)} (\delta^p_q {}^1K^r_s + \delta^r_s {}^1 K^p_q + {}^2 V^{p,r}_{q,s}(N-1) )
\end{equation}
we define an operator $W^{p,r}_{q,s}={}^2K^{p,r}_{q,s} - {}^2 K^{p,r}_{s,q}$, which then leads to an expression for the total ACSE equation as \cite{Mazziotti2006_prl, Mazziotti2007_pra}:
\begin{equation}
    \begin{split}
    {}^2 {A}^{i,k}_{j,l} &= \sum_{p,q} {}^2 D^{p,q}_{i,k} W^{p,q}_{j,l} - {}^2 D^{p,q}_{j,l} W^{p,q}_{i,k} \\
    &+ \sum_{pqr} {}^3 D^{p,r,k}_{j,l,q} W^{p,r}_{i,q}
    - {}^3 D^{p,r,i}_{j,l,q} W^{p,r}_{k,q}
    - {}^3 D^{i,k,p}_{r,q,j} W^{p,l}_{r,q}
    + {}^3 D^{i,k,p}_{r,q,l} W^{p,j}_{r,q}.
\end{split}
\end{equation}

Notably, this expression involves the 3-RDM, which can be reconstructed from its cumulant expansion~\cite{Mazziotti1998}:
\begin{equation}
{}^3 D = {}^1 D \wedge {}^1 D \wedge {}^1D + 3~{}^2 \Delta \wedge {}^1 D + ^3 \Delta
\end{equation}
Here, the wedge product denotes Grassmannian operator, combining antisymmetric permutations of upper and lower indices, ${}^n \Delta$ represents the $n$-th order reduced cumulant matrix, and we assume that ${}^3 \Delta=0$.

\subsection{Circuit Implementations}
Once the ${}^2 A$ operator is obtained for each step, we use a threshold to truncate the operator, and at each step add only one or two additional fermionic terms. We list some examples operators which were present for the 3- and 4- qubit cases in the Supplemental Material. \tc{As mentioned in the main text, }the circuits are constructed by expressing $e^{\epsilon_n A_n}$ as a first order trotterization, resulting in products of exponentials Pauli strings which can be realized generally with CNOT gates and single-qubit rotations.  In some instances we see a reduction in the number of two-qubit gates by using the following single-qubit identity:
\begin{equation}
e^{i\pi} U^\dagger \sigma_j U  =
    \begin{cases}
        \sigma_x  & \text{if $j$=$x$} \\
        \sigma_y  & \text{if $j$=$z$} \\
        \sigma_z  & \text{if $j$=$y$} \\
    \end{cases}
\end{equation}
where $U= S^\dagger H S $. This can just as easily applied to exponential transformations as well, and with this, we can transform an operator such as $e^{\alpha (X_1 X_2+ Y_1 Y_2)}$, which is expressed in 3 or 4 CNOT gates, to $U^\dagger e^{-\alpha (X_1 X_2 + Z_1 Z_2)}U$ which can be expressed with only 2 CNOT gates. In general, we utilize straightforward concatenation techniques which possibly reduced the CNOT gates while preserving the connectivity of the device (which is linear).

\section{Error Mitigation Methods}
To directly mitigate the effects of noise on the quantum computer, we use a variety of techniques in addition to the ones listed in the main text (limit-preserving correction and the purification of the 2-RDM).

\subsection{Number Preserving Projection to Diagonal Elements of the 2-RDM}
The most effective error correction comes by filtering diagonal elements of the 2-RDM, of the form ${}^2 D^{p,q}_{p,q}$, so that the number operator is preserved. Because these elements commute with single-qubit measurements that are performed, they can be filtered according to the measurement result. Counts that have differing values of $N$ or $S_z$ are rejected, and so we are filtered to a set of RDMs with the proper trace and projected spin properties (i.e., ${\rm Tr~}{}^2 D = N(N-1)$). While heavily erroneous off-diagonal elements can also lead to non-physical eigenvalues\cite{Arute2020}, correcting for these in the 2-RDM case is not straightforward and likely would not reduce the overall errors.

\subsection{Measurement Correction of Prepared States (SPAM)}
Finally, the state preparation and measurement, which involves preparing all possible quantum states for some qubit space, and constructing a transition matrix with the associated inverse, was utilized to mitigate measurement errors. We applied this to local qubits, and so did not correct for correlated measurement errors. This procedure has been documented in many places~\cite{Breuer2007, Govia2020} and can be implemented through \textsc{qiskit}.

\section{Symmetry and the ACSE}
Given a symmetry operator $\hat{S}$ (where $[\hat{S},\hat{H}]=0$) utilizing the qACSE method leads to natural advantages in terms of the generated ansatz and preserving the symmetry subspace. In particular, we can readily see that any symmetry of the system is not violated throughout the qACSE iterations.

Take a particular iteration of the ACSE algorithm for a generic quantum system, where we are in a single symmetry state $s_0$ of the symmetry $\hat{S}$. Then, we can write our Hamiltonian and state as:
\begin{align}
    \hat{H} &= \sum_{i,s} H_{i,s} |i,s\rangle \langle i,s | \\
    |\Psi_n \rangle &= \sum_{k} d_n^{k,s_0} |k,s_0 \rangle .
\end{align}
where $|i,s\rangle$ represents a state $i$ within the symmetry subspace of $s$ and $H_{i,s}$ and $d_{k,s_0}$ are coefficients. In this formulation, elements of $\hat{A}$ can be found as:
\begin{align}
    A_n^{\alpha u;\beta v} &= \langle \psi_n | [\hat{M}^{\alpha,u}_{\beta,v},\hat{H}]| \psi_n \rangle \\
    &= \sum_{k,j} \sum_{i,s} d_n^{k,s_0} d_n^* {}^{j,s_0} \langle k,s_0 |[ |\alpha,u \rangle \langle \beta,v|, \hat{H}] |j,s_0\rangle  \\
    &= \sum_{k,j} d_n^{k,s_0} d_n^* {}^{j,s_0} (H_{j,s_0} \delta^{k,s_0}_{\alpha,u} \delta^{\beta,v}_{j,s_0} - H_{k,s_0} \delta^{k,s_0}_{\alpha,u} \delta^{\beta,v}_{j,s_0})
\end{align}
where $\hat{M}^{\alpha,u}_{\beta,v}$ represents a measurement operator between two basis elements $|i,t \rangle $ and $|j,u\rangle$. This expression is clearly nonzero only if $u=s_0$ and $v=s_0$, and so at each step we will preserve whatever symmetry subspace we are in, which means that throughout the algorithm the symmetry state of system is preserved.

Note that this only applies if we are in the symmetry basis of $\hat{S}$. Practically, this is not always the case. For instance, the standard second quantized representation for a fermionic simulation uses Slater determinants, which commute with the number and projected spin operators (Slater determinants). While using the entire ${}^2 A$ operator will preserve all symmetries, using a truncated ${}^2 A$ operator can lead to symmetry violations in the total spin, but not projected spin or number operator. Through a spin adapted operator basis, this can be easily overcome.

\bibliography{
    b0_vqe.bib,
    b1_qc_err_mit.bib,
    b2_acse.bib,
    b3_benzyne.bib,
    b4_qc_misc.bib,
    b5_software.bib,
    b6_rdm_methods.bib,
    b7_cse.bib,
    b8_supp.bib,
    QACSE-Benzynes-ACSE
    }
\end{document}